\documentclass[pra,twocolumn,showpacs,preprintnumbers,amsmath,amssymb,floatfix]{revtex4}
\usepackage{epsfig}
\usepackage{graphicx}
\usepackage{amssymb}
\usepackage{color}

\newcommand{\ket}[1]{\left\vert#1\right\rangle}

\newcommand{\Ham}{\mathcal H}
\newcommand{\beq}{\begin{equation}}
\newcommand{\eeq}{\end{equation}}
\newcommand{\bea}{\begin{eqnarray}}
\newcommand{\eea}{\end{eqnarray}}

\begin{document}

\title{Chopped random basis quantum optimization}

\author{Tommaso Caneva}
\author{Tommaso Calarco}
\author{Simone Montangero}

\affiliation{
Institut f\"ur Quanteninformationsverarbeitung,
  Universit\"at Ulm, D-89069 Ulm, Germany}

\date{\today}

\begin{abstract}
In this work we describe in detail the \emph{Chopped RAndom Basis}
(CRAB) optimal control technique recently introduced 
to optimize t-DMRG simulations~\cite{Doria_10:preprint}. Here we study the efficiency 
of this control technique in optimizing different quantum processes and we
show that in the considered cases we obtain results equivalent to
those obtained via different optimal control methods while using less
resources. We propose the CRAB optimization as a general and
versatile optimal control technique. 
\end{abstract}

\maketitle


Realizing artificial, controllable quantum systems has represented
one of the most promising challenge in physics for the
last thirty years~\cite{Brif_NJP10}.
On one side such systems could unveil unexplored features of Nature, when employed 
as universal quantum simulators~\cite{Feynman_IJTP82}; on the other side this technology 
could be exploited to realize a new generation of extremely powerful
devices, like quantum computers~\cite{Lloyd_NAT00}. Along with the impressive
progress marked recently in the construction
of tunable quantum systems~\cite{Bloch_RMP08,Roati_NAT08}, there is a renewed and 
increasing interest in quantum optimal control (OC) theory, the study of the optimization
techniques aimed at improving the outcome of a quantum
process~\cite{Brif_NJP10}. Indeed OC can prove to be  
crucial under several respects for the development of quantum
devices: first, it can be generally employed to speed up a quantum
process to make it less prone to decoherence or noise effects 
induced by the unavoidable interaction with the external environment. 
Second, considering a realistic experimental setup in which just few parameters are 
tunable or, in the most difficult situations, only partially tunable, OC can
provide an answer about the optimal use of the available resources.

Traditionally OC has been exploited in atomic and molecular 
physics~\cite{Peirce_PRA88,Calarco_PRA04,Khaneja_JMR05}. 
More recently, with the advent of quantum information,
the requirement of accurate control of quantum systems 
has become unavoidable to build quantum information 
processors~\cite{Bose_PRL03, Montangero_PRL07, Krotov:book, 
Carlini_PRL06, Rezakhani_PRL09,Verstraete_NAT10, Sporl_PRA07}. 
However, the above mentioned methods often result in optimal driving fields
that require a level of tunability incompatible with current 
experimental capabilities and in general, the calculation of the
optimal fields requires an exact description of the system (either analytical or
numerical). The field of application of these methods is severely
limited also by the need to have access to huge amount of information
about the system, e.g. computing gradients of the control fields, expectation
values of observables as a function of time. Moreover, standard OC
algorithms define a set of Euler-Lagrange equations that have to be
solved to find the optimal control pulse~\cite{Brif_NJP10}, where the
equation for the correction to the driving field is highly dependent
on the constraints imposed on the system and on the figures of merit
considered. This implies that considering different figures of merit
and/or constraints on the system needs a redefinition of the
corresponding Euler-Lagrange equations, hindering a straightforward
adaptation of the optimization procedure to different situations. 

In this work we discuss in detail the \emph{Chopped RAndom Basis} (CRAB)
technique,  an optimization method 
directed to overcome these difficulties and
 already introduced in~\cite{Doria_10:preprint}.
The CRAB optimization is based on the definition of a truncated 
randomized basis of functions for the control fields that recast the problem from a
functional minimization to a multi-variable function minimization
that can be performed, for example, via a direct-search method. As
shown in the following, the CRAB optimization flexibility 
allows to construct OC pulses just exploiting 
the available resources. Indeed, different 
figures of merit and constraints can be easily considered without 
any complications. Another appealing characteristic of CRAB  
is its compatibility with t-DMRG techniques: 
this feature indeed significantly enlarges the class of controllable 
systems~\cite{Doria_10:preprint}, from few-body or exactly solvable to
general many-body quantum systems with ``moderate'' degree of entanglement
generated during the dynamics~\cite{Schollwock_RMP06}. This is, to the best
of our knowledge, the unique OC algorithm that can be applied in such
vast setting. Finally, it can be straightforward applied also in a closed-loop
optimization experiment, where the simulation of the system under
study is replaced with the experiments itself. 

Here we analyze the CRAB optimization as a possible general 
OC algorithm to be used also in a standard context
(solvable and/or few body systems) as a valid alternative tool with respect to
standard OC methods to find optimal control fields. 
Indeed, recently optimization methods based on the expansion over
a particular function basis have shown to be 
effective~\cite{Romero_PRA07, Jiang_PRA09, Galve_EPJ10, DiFranco_PRA10}.
In particular, a similar approach has been proved to be mathematically  
convergent and consistent~\cite{Jr-ShinLi_JCP09,Ruths_11:preprint}.
On top of that, some theoretical analysis
over control landscapes suggests that, at least in the absence of
constraints, the figure of merit landscape might be smooth enough to 
allow for simple optimization procedures to
work~\cite{Rabitz_SCI04, Rabitz_PRA06}. Here, we show that
indeed a convenient choice of the function basis driven by 
physical or geometrical arguments is enough to obtain
optimal driving fields. However, in the cases where no physical 
intuition drives the choice of the function basis, the CRAB algorithm allows 
to find the optimal driving fields where a simple ansatz would fail.
Moreover, a comparison between the results of CRAB with and without 
a physically driven choice of the basis, as well as previous results 
obtained using different optimal control algorithms (Krotov's
algorithm), show comparable performances~\cite{Caneva_10:preprint}.

The structure of the paper is the following: in Sec.~\ref{crab:sec} the CRAB
optimization is described; in Sec.~\ref{joseph_crab:sec} 
it is applied to a paradigmatic quantum control problem, 
the state transformation of two coupled qubits, to show its potential. Then we
compare the results obtained via CRAB optimization in more complex cases already
present in literature~\cite{Caneva_PRL09,Caneva_10:preprint}: 
in Sec.~\ref{lmg_crab:sec} the method is employed to 
control the quantum phase transition evolution of the
Lipkin-Meshkov-Glick (LMG) model; and in Sec.~\ref{transfer_crab:sec}
we optimize the transfer of a state along a spin chain. 
Finally, in Sec.~\ref{lmg_entanglement:sec} the optimization 
is exploited to maximize the final entanglement entropy of the final
state in the LMG model;
and in Sec.~\ref{lin_antiad:sec} a comparison between adiabatic and
optimized processes is proposed.

\section{CRAB optimization}\label{crab:sec}
%
The optimization problem we are dealing with is defined as follows:
given a Hamiltonian $H$ acting on a Hilbert space $\mathcal{H} = \mathbb{C}^N$, 
depending on a set of time-dependent driving fields 
$\vec{\Gamma}(t)$, we search for the optimal transformation to drive,
in time $T$, an initial state $|\psi _0\rangle \in \mathcal{H}$ into a different
one (target state) $|\psi_G\rangle  \in \mathcal{H}$ with some desired 
properties expressed by  a cost function $f(|\psi_G\rangle)$ we 
want to minimize~\footnote{The generalization of the problem to the optimization of an 
overall unitary transformation is straightforward, averaging over the
contributions of a complete set of basis of the Hilbert space
$\mathcal{H}$.}. In addition, constraints might be present on 
the driving fields, e.g. to match
experimental conditions: They can be expressed usually as a
function of the driving fields $\mathcal{C}_i(\vec \Gamma(t))$.  
Typical scenarios and corresponding cost functions and constraints 
are:
\begin{enumerate}
\item{The goal is the preparation of a well-defined quantum state $|\psi _G\rangle$
  with high accuracy for which a convenient cost function is the infidelity,
\begin{eqnarray} 
f_1(|\psi (T)\rangle)\equiv\mathcal{I}(T)=1-|\langle\psi (T)|\psi _G
\rangle|^2.
\label{effe}
\end{eqnarray}}
\item{The target state is the unknown ground state of a
Hamiltonian $H_p$. The cost function is then given by
the final system energy,
\begin{eqnarray} 
f_2(|\psi (T)\rangle)\equiv E_f(T)=\langle \psi (T)|H_p|\psi (T)\rangle.
\label{e_fin_crab:eq}
\end{eqnarray}}
\item{The target is some property or condition that many 
states can satisfy, like for example, in the production of highly 
entangled states. In this case the cost function is 
simply defined as%
\begin{eqnarray} 
f_3(|\psi (T)\rangle)\equiv-S(|\psi (T)\rangle),
\label{ent_crab:eq}
\end{eqnarray}
where $S(|\psi \rangle)$ is a convenient measure of the entanglement of the state
$|\psi\rangle$.}
\item{A constraint is present on the power of the driving
  fields, that is, the solution should minimize also the fluences
\begin{equation} 
\mathcal{C}_i= \int |\Gamma_i(t)|^2 dt
\label{fluence}
\end{equation} }
\item{A limited bandwidth is allowed for the driving fields: 
  below we show how this is already embedded in the algorithm and
  is not necessary to consider it as an additional explicit constraint.}
\item{The initial state or the driving fields are known within a
  given uncertainty $\epsilon$. In this case, the cost function
  can be defined as an average other all possible outcomes compatible
  with that uncertainty, as for example:
\begin{equation} 
f_4= \int f(|\psi (T,\epsilon)\rangle) d\epsilon.
\label{uncert}
\end{equation}
}
\end{enumerate}

All of the aforementioned optimization problems are then recast in the problem
of solving the Sch\"odinger equation 
(from now on we assume $\hbar=1$)
\begin{eqnarray}
 i\frac{d}{dt}|\psi (t)\rangle = H[\vec{\Gamma}(t)]|\psi (t)\rangle ,
\label{schr_eq}
\end{eqnarray}
with boundary condition $|\psi _i\rangle=|\psi (0)\rangle$, while minimizing the 
cost function 
\begin{eqnarray}
\mathcal{F} = \alpha f + \sum_i \beta_i
\mathcal{C}_i(\Gamma(t)),
\label{costF}
\end{eqnarray}
where the coefficients $\alpha$ and $\beta_i$ allow for a proper 
weighting of the different contributions (the $\beta$s play the role of 
Lagrange multipliers) and $f$ is the chosen cost function. 

To perform such an optimization, the CRAB algorithm starts from 
an initial pulse guess $\Gamma _j^0 (t)$ and then looks 
for the best correction of the form
\begin{eqnarray}
\Gamma _j^{\rm CRAB}(t)=\Gamma _j^0(t)\cdot g_j(t).
\end{eqnarray}
The functions $g_j(t)$ are expanded in a simple form in some
function basis characterized by some parameters $\vec{\Omega}_j$
(Fourier space, Lagrange polynomials, etc.):
$g_j=\sum_k\,c_j^k\,\hat g_j^k(\Omega_j^k)$.
The two key ingredients of the CRAB optimization are
that the function space is truncated to some finite number of
components $N_c$ ($k=1, \dots, N_c$) and that the corresponding basis
functions are ``randomized'' to enhance the algorithm convergence, 
i.e. $\hat g_j^k \rightarrow \hat g_j^k(\Omega_j^k (1+r_j^k))$ where $r_j^k$ is a
random number. Indeed, this last choice breaks the orthonormalization
of the functions $ g_j^k$, however as we show in the following, it allows
for an improved convergence of the algorithm as it enlarge the subspace
of functions explored by the algorithm while keeping constant the
number of optimization parameters.

The optimization problem is then reformulated as the extremization of the 
multivariable cost function $\mathcal{F}(T,\vec{c}_j)$,
which can be numerically approached with a suitable method, e.g., steepest
descent, conjugate gradient or direct search methods~\cite{NumericalRecipes}.
Hereafter we use the last option, which is the simplest one and easily
compatible with any technique employed to solve the dynamics induced by $H[\Gamma (t)]$
(either exact solution of the Eq.~(\ref{schr_eq}) or approximate solution with time dependent 
DMRG~\cite{Schollwock_RMP06}). This choice also gives another advantage with
respect to other OC methods where gradients and functional 
derivatives have to be computed, increasing the complexity of the optimization procedure.  
 

\begin{figure}[t]
\epsfig{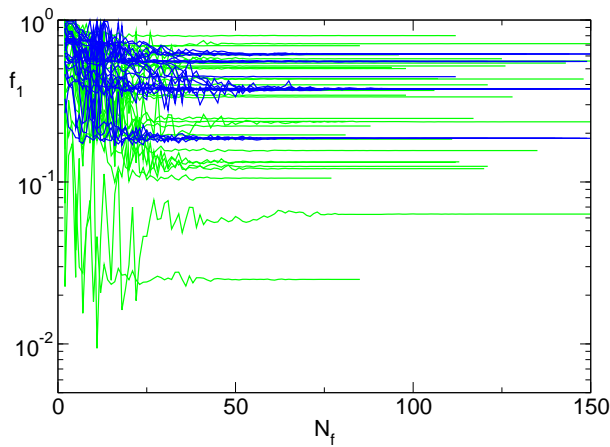}
\caption{Infidelity $f_1$ of the final state 
as a function of number of calls to the optimization
algorithm $N_f$ for two capacitively Josephson charge qubits with
principal harmonics (dark grey [blue] line) and randomized frequencies (light grey [green] line),
for the goal state $|\psi_G^1\rangle$ and $N_c=2$ for thirty different
random instances.}
\label{err}
\end{figure}

As an example, in the following problems, we focus on the case of a single control parameter $\Gamma (t)$
and we choose to work in the Fourier basis. The optimal pulse can 
then be written as
\begin{eqnarray}
g(t)= 1 + 
\frac{\left(\sum _{n=1}^{N_c}A_n \sin(\omega _n t) + B_n \cos(\omega_nt) \right)} {\lambda (t)},
\label{field_crab:eq}
\end{eqnarray}
where $\lambda (t)$ is a time dependent function enforcing the boundary conditions
(i.e. $\lambda (t)\rightarrow\infty$ for $t\rightarrow 0$ and for
$t\rightarrow T$). The function $\Gamma^{\rm CRAB}(t)$ is 
fixed by selecting the optimization parameters $ \vec{A},\vec{B}$ and 
$\vec{\omega}$, with $N_c$ the dimension of each vector.
In conclusion, given a fixed total evolution time
$T$, the cost function is clearly just a function of the control parameters,
\begin{eqnarray} 
\mathcal{F}=\mathcal{F}^{\rm CRAB}(\vec{A},\vec{B},\vec{\omega}).
\label{inf_crab:eq}
\end{eqnarray}
The optimization problem is reduced to the minimization of  
$\mathcal{F}^{\rm CRAB}(\vec{A},\vec{B},\vec{\omega})$ as a function of 
$3\times N_c$ variables. As mentioned before, however, the space of
the variables can be reduced even more: although in principle
the frequencies $\vec{\omega}$ can be considered free variables
it is often convenient to keep them fixed and to perform the minimization just
with respect to $\vec{A}$ and $\vec{B}$. Indeed as shown in our analysis
this is sufficient to obtain good results. In this approach 
we need then a criterion to select the $\vec{\omega}$'s.
When we have no available information about the typical energy scales of the system
under consideration, the frequencies are picked \emph{randomly} around
principal harmonics: $\omega_ k=2\pi k(1+r_k)/T$, 
with $r_k$ random numbers with flat distribution in the interval $[-0.5,0.5]$ and 
$k=1,...,N_c$. Viceversa when the physical details of the model are known,
clearly one can exploit this information to select the relevant
frequencies, as shown in the following sections.

\section{Two-qubits optimization}\label{joseph_crab:sec} %

\begin{figure}[t]
\epsfig{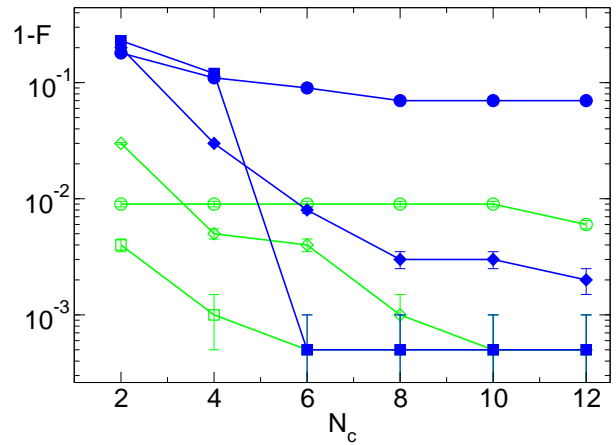}
\caption{Optimized infidelity $f$ as a function of the number of
optimization parameters $N_c$  
with principal harmonics (dark grey [blue], full symbols) and randomized frequencies
(light grey [green], empty symbols)
for different target states
$|\psi_G^1\rangle$ (circle), $|\psi_G^2\rangle$ (squares), 
$|\psi_G^3\rangle$ (diamonds).}
\label{minerr}
\end{figure}

In this section we apply the CRAB optimization to a paradigmatic
problem in quantum information theory and control: we search for the
optimal way to perform a state transformation of a two-qubit system,
in particular we consider two capacitively coupled Josephson charge
qubits, even though the following analysis can be easily adapted to
different qubit implementations. The Hamiltonian of the $i$-th qubit 
is defined as~\cite{Mahklin_RMP01,Wendin_06:inc}
\begin{equation*}
    \Ham_i  = E_C \sigma_z^i + E_J \sigma_x^i
\label{hamloc}
\end{equation*}
where the $\sigma$s are Pauli matrices, 
$E_C$ is the charging energy and $E_J$ is the
Josephson energy and $i=1,2$. For capacitive coupled qubits, 
the interaction Hamiltonian reads
\begin{equation*}
    \Ham_I\! =  E_{cc} \sigma_z^1 \sigma_z^2, 
\label{hamintc}
\end{equation*}
where $E_{cc}$ is the charging energy associated to the Coulomb
interaction between the qubits. Hereafter we set $E_J/E_C=-1$, 
while the coupling will be the driving field 
$E_{cc}(t)/E_C=\Gamma(t)$ we use to optimize the transformation. 
We will consider as initial state the state with no excess Cooper
pairs $|\psi_0\rangle = \ket{00}$, and our goal states will be three
different state with different properties: the reversed separable 
state $|\psi_G^1\rangle = \ket{11}$, the homogeneous superposition state 
$|\psi_G^2\rangle =  \frac{1}{2} \sum_{i,j} \ket{i,j}$, and 
the maximally entangled Bell state 
$|\psi_G^3\rangle = \frac{1}{\sqrt{2}}(\ket{00}+\ket{11})$. 
Note that due to the fact that only the coupling is controlled, all
three states are not trivial to achieve.
We set the total time of the transformation to the somehow arbitrary time scale 
$T=\pi/E_J$ and we perform a CRAB optimization using the truncated expansion of
the function $g(t)$ given in Eq.~(\ref{field_crab:eq}), with a constant
initial guess for the driving field $\Gamma^0(t)=\Gamma(0)=1$. We
considered an additional constraint on the fluence of the control
field, thus the resulting cost function is defined as 
\begin{figure}
\epsfig{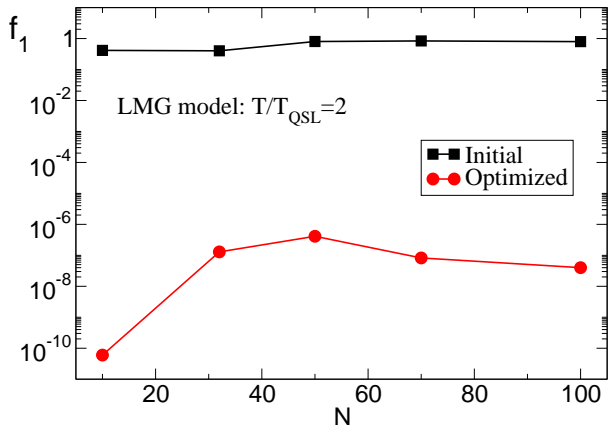}
\caption{(Color online) Infidelity as a function of the size in the
  LMG model. Squares represents the data before the optimization,
  circles the data after the optimization with CRAB. 
}
\label{crab_lmg_inf_vs_size_fig}
\end{figure}
\begin{equation}
\mathcal{F} =  f_1 + 0.1 \, \mathcal{C}_1(\Gamma(t)),
\end{equation}
where $f_1$ and $ \mathcal{C}_1$ are given by equations~(\ref{effe})
and~(\ref{fluence}) respectively. 
Here we are interested in studying the effect of the randomness
introduced in the frequencies of the expansion~(\ref{field_crab:eq}),
thus we optimize both in the case of random $r_k$ and with $r_k=0$. To
perform a fair comparison, we ran the optimization in both cases with
the same maximum number of calls $N_f \sim 30.000$ to the function $\mathcal{F}$,
which fixes the simulation complexity. Indeed, in the first case we repeated the
optimization for thirty different $r_k$ random configurations (with a
single $A_k,B_k$ random starting point), while in
the second case the optimization was repeated over thirty 
initial random $A_k,B_k$ configurations. A typical result is shown in 
Fig.~\ref{err} for $N_c=2$ and $|\psi_G^1\rangle$: it clearly shows
that for the case of randomized $\omega_k$ the optimization is highly
improved (notice the logarithmic scale). A more systematic 
comparison is shown in Fig.~\ref{minerr} where the best results are 
plotted against the number of optimization
parameters $N_c$ for the three target states $|\psi_G^i\rangle$: 
in all cases the randomization of the frequencies improves the
convergences to higher fidelities up to the simulation
error. In particular, in one case, the final result without 
randomization is very far from being satisfactory as the final 
fidelity is of the order of ten percent, resulting in a 
very poor state transformation. On the contrary,
using the randomized frequencies we were able to find optimal
pulses to obtain fidelities below one percent -- values that are 
comparable, in most cases, with experimental errors.
%

\section{Lipkin-Meshkov-Glick model}\label{lmg_crab:sec} %
%
\begin{figure}
\epsfig{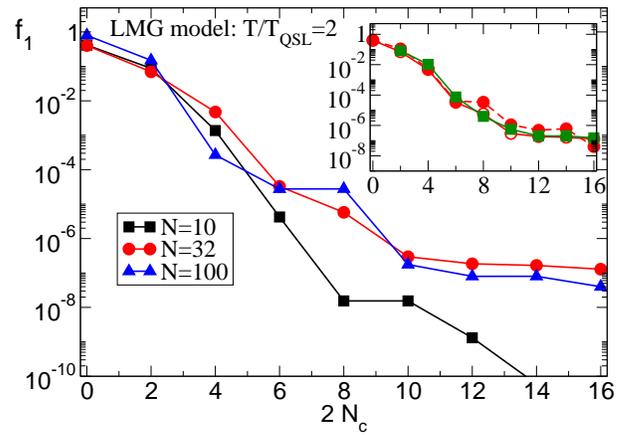}
\caption{(Color online) Infidelity as a function of the number of
control parameters for different sizes in the LMG model. 
The total evolution time is $T=2T_{\rm QSL}=2 \pi/ \Delta$. Inset: infidelity as a 
function of the number of parameters for a single size $N=32$: 
comparison between data optimized using as cost function the infidelity
(empty circles) and the final energy (full circles). Green squares
represent the results with randomized frequencies.}
\label{crab_lmg_inf_vs_npar_fig}
\end{figure}
The Lipkin-Meshkov-Glick (LMG) model is the paradigm of a system 
with long range interaction (infinite in the thermodynamical limit). 
The Hamiltonian 
is written as~\cite{Lipkin_NP65,Botet_PRB83}:
\begin{eqnarray}
H=-\frac{J}{N}\sum _{i<j}(\sigma_i ^x \sigma_j ^x +\gamma \sigma_i ^y \sigma_j ^y)
    -\Gamma (t)\sum _{i}^N \sigma_i ^z ,
\label{LMG_ham:eq}
\end{eqnarray}
where $J$ is the uniform spin-spin interaction (we set $J=1$ in the following), 
$N$ is the number of spins in the system,   
$\Gamma$ is the transverse field and $\sigma ^{\alpha}_i$ are the Pauli matrices.
By introducing the total spin operator $\mathcal{S}_{\alpha}=\sum_i
\sigma ^{\alpha}_i/2$,  Eq.~(\ref{LMG_ham:eq}) can be rewritten,
 apart from an additive constant, as
$H=-\frac{1}{N}[\mathcal{S}_x^{2} +\gamma \mathcal{S}_y^{2}]
-\Gamma \mathcal{S}_z$. 
The Hamiltonian hence commutes with $\mathcal{S}^2$ and does not
couple states having a 
different parity in the number of spins pointing in the magnetic field
direction: $[H,\mathcal{S}^2]=0$ and $[H, \prod _i \sigma_i ^z]=0$. In the isotropic 
case $\gamma=1$, also the $z$-component of $\mathcal{\vec{S}}$ is conserved,
$[H,\mathcal{S}_z]=0$. 
In the thermodynamical limit the LMG model undergoes a second order quantum phase 
transition at $\Gamma _c =1$ from a paramagnet ($\Gamma > 1$) to a ferromagnet
($\Gamma < 1$). The phase transition is 
characterized by mean-field critical exponents~\cite{Botet_PRB83}. 
The phase transitions dramatically affects the dynamical behavior 
of quantum systems: As discussed in more detail in Sec.~\ref{lin_antiad:sec}, 
the gap closure at the critical point promotes dynamical excitations, 
preventing adiabatic evolutions whenever the 
adiabaticity condition $T\gg \Delta^{-1}$ is not fulfilled, 
where $T$ is the total evolution time and 
$\Delta$ the minimum spectral gap~\cite{Messiah:book,Sachdev:book,Zurek_PRL05,Polkovnikov_NAT08,Pellegrini_PRB08,
Deng_EPL08,DeGrandi_09:preprint,Divakaran_09:preprint,Dziarmaga_AP10,Polkovnikov_10:preprint}. 
Following Ref.~\cite{Caneva_10:preprint}, we employ the CRAB optimization 
to drastically reduce the residual density of defects present in the
system in a strongly non adiabatic dynamics, drastically reducing the
time needed to connect the ground state in 
one phase with the ground state of the other phase with respect to adiabatic
non-optimized strategies.
We chose as initial state the ground state (gs) of $H[\Gamma(t)]$ at
$\Gamma _i\gg 1$,  i.e. the state in which all the spins
are polarized along the positive $z$-axis (paramagnetic phase).
As target state we chose the gs of $H[\Gamma=0]$ (ferromagnetic phase).
We focused our attention on the case $\gamma=0$, representative of the
class $\gamma<1$ (for $\gamma=1$ the dynamics is trivial due to the 
symmetry of $H$)~\cite{Caneva_PRB08}.  
For this model indeed a lot of physical information is available:
the gap between the ground state and the first
excited state closes polynomially with the size at the 
critical point~\cite{Botet_PRB83}, 
$\Delta \sim N^{-1/3}$. 
Furthermore it has been recently demonstrated that the minimum time required to 
obtain a perfect conversion between the initial and the final state here considered, 
the so called \emph{quantum speed limit}, is given 
by $T_{\rm QSL}=\pi/\Delta$~\cite{Giovannetti_PRA03,Caneva_10:preprint}. 
In order to test the performance of CRAB, we fixed the total
evolution time above this threshold, at $T=2T_{\rm QSL}$, in a regime in which 
in principle it is possible to produce an arbitrarily small infidelity with 
optimized evolutions.
\begin{figure}
\epsfig{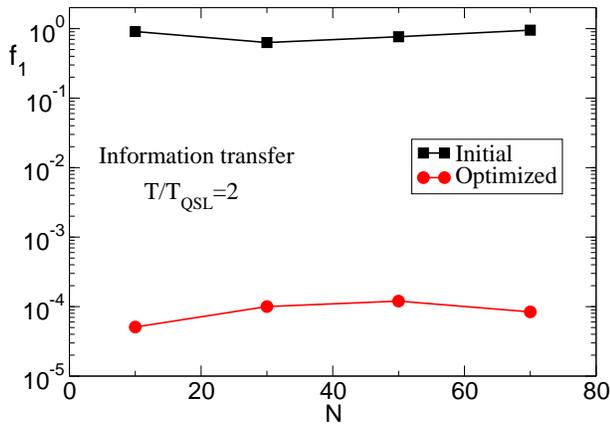}
\caption{(Color online) Infidelity as a function of the size in the
  transfer state problem.  
Squares represents the data before the optimization, circles the data
after the optimization with CRAB.}
\label{crab_chain_inf_vs_n_fig}
\end{figure}

The results of our simulations for the LMG model are summarized in 
Fig.~\ref{crab_lmg_inf_vs_size_fig} and Fig.~\ref{crab_lmg_inf_vs_npar_fig}; the
data shown in the two pictures (with the only exception of the inset of 
Fig.~\ref{crab_lmg_inf_vs_npar_fig} as explained in the following)
have been produced assuming Eq.~(\ref{field_crab:eq}) as control field
and the infidelity 
as cost function to minimize.
In Fig.~\ref{crab_lmg_inf_vs_size_fig} we plotted the infidelity as a function
of the size $N$, before the optimization for a linear driving field 
$\Gamma^{0} (t)\propto t/T$
(squares), and after the optimization with CRAB (circles): for each size
we have been able to produce an infidelity below $10^{-6}$ starting from 
an infidelity of order $\mathcal{O}(1)$. In particular the data have been produced
by minimizing Eq.~(\ref{inf_crab:eq}) with respect to $\vec{A}$ and $\vec{B}$,
while keeping $\vec{\omega}$ fixed, for a total of $2\times N_c=16$ parameters.
In this case the frequencies $\vec{\omega}$ have been chosen by exploiting
the physical information available. We chose the frequencies 
equal to the minimum spectral gap  
$\omega _1=2\pi /T=2\pi/2T_{\rm QSL}=\Delta $ and we considered 
the main harmonics $\omega _k=k\omega _1$ for $k$ up to $N_c$.
In Fig.~\ref{crab_lmg_inf_vs_npar_fig} we plot the infidelity as a 
function of number of parameters employed to build the optimal field of 
Eq.~(\ref{field_crab:eq}) -- adding a frequency $\omega_ k$ corresponds to 
add two parameters, $A_k$ and $B_k$. First it can be noticed that 5 
harmonics are sufficient to reach the best optimization result, 
$\mathcal{I}\sim 10^{-6}$; however with only 3 harmonics the infidelity is already
of order $10^{-4}$, of the order of the required threshold for
fault-tolerant quantum computation. 
Considering the implementation of an optimal pulse in an 
NMR or quantum optics experiment, the gain with respect to other OC
methods providing 
a totally arbitrary $\Gamma _{\rm opt}(t)$ is evident. The second interesting 
feature is that the behavior of the infidelity in Fig.~\ref{crab_lmg_inf_vs_npar_fig} 
is approximately independent of the size (for the smallest system
considered, $N=10$, 
finite size effects are more evident): this confirms the intuition that the most 
relevant energy scale for the LMG model is given by the minimum 
spectral gap.\\
Finally, in order to verify the independence of the optimization from the
knowledge of the target state, we repeated the simulations
\begin{figure}[t]
\epsfig{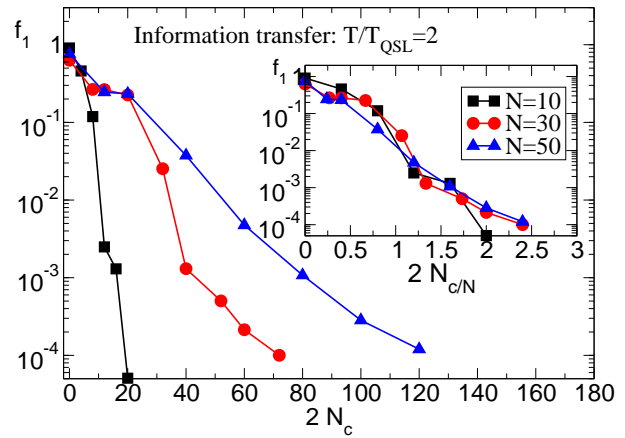}
\caption{(Color online) Infidelity as a function of the number of
  control parameters for different  
sizes in the in the state transfer model. The total evolution time
$T=2T_{\rm QSL}$. Inset: infidelity as a function of the number 
of parameters divided by the size.}
\label{crab_chain_inf_vs_npar_fig}
\end{figure}
assuming as a cost function the final energy $E_f(T)$ of Eq.~(\ref{e_fin_crab:eq}).
In the inset of Fig.~\ref{crab_lmg_inf_vs_npar_fig} we compare the infidelity
of the data optimized using as cost function the infidelity itself (empty circles)
and the final energy (full circles), for a specific size of the system $N=32$
and for different number of control parameters: as shown in the 
picture the agreement is very good. We also repeated the optimization
using randomized frequencies, obtaining the same results as before. 
Thus, also in the case where the chosen frequencies are optimal,
introducing randomness does not prevent the optimization to
work. On the contrary, if one has no access to any information on the
system, the randomization does not prevent to reach the same optimal result.

\section{State transfer along a spin chain}\label{transfer_crab:sec}
\begin{figure}
\epsfig{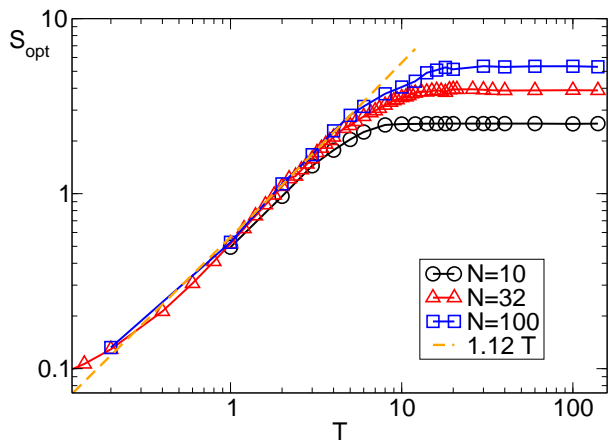}
\caption{(Color online) Final entanglement entropy as a function of 
the total evolution time $T$ for $N=10,32,100$ of one of two equal
bipartitions of the system.
The time is measured in units of $J^{-1}$.}
\label{lmg_crab_ent_vs_T_fig} 
\end{figure}
In this section we study the optimization of a model representing a
possible implementation 
of a quantum bus. The model consists in a chain of spins coupled via
uniform nearest-neighbor (n.n.) 
interaction; by acting with an external, parabolic magnetic field it is possible
to transfer a quantum state along the 
chain~\cite{Bose_PRL03,Caneva_PRL09,Murphy_PRA10}.  
In particular we follow the lines of Ref.~[\onlinecite{Balachandran_PRA08,Murphy_PRA10}].
The Hamiltonian of the system is 
\begin{eqnarray}
 H(t)=-\frac{J}{2}\sum _{n=1} ^{N-1}\vec{\sigma}_n\cdot\vec{\sigma}_{n+1}+\sum _{n=1} ^{N}B _n(t)\sigma ^z _n,
\end{eqnarray}
where $N$ is the number of spins in the chain, $\vec{\sigma}_n$
represents the Pauli $n-$th-spin operator, 
$J$ is the uniform n.n. interaction (we set $J=1$ in our simulations),
and $B _n(t)$ is the tunable magnetic field along the $z$-direction.
In particular we considered a parabolic magnetic field tunable in position and 
strength~\cite{Murphy_PRA10},
\begin{eqnarray}
 B_n(t)=C(t)(x _n-d(t))^2,
\end{eqnarray}
where $d(t)$ is the position of the potential minimum along the chain,
$x_n$ is the position of  
the $n-$th spin, and $C(t)$ is the instantaneous curvature of the field.
Far from the minimum, the spins are forced by the magnetic field to be
aligned along the $z$-axis 
irrespective of their mutual interaction; instead close to the minimum,
the n.n. coupling prevails 
and can be exploited to transfer the information (i.e. the state) from
one site to the next. 
The Hamiltonian commutes with the total magnetic field along the $z$-direction, 
$[H(t),\sum _{n=1}^N\sigma _n^z]=0$, so that the dynamics occurs in a subspace whose 
dimension grows just linearly with the size $N$ of the system.
We chose to work in the subspace $\langle\sum _{n=1}^N\sigma _n^z\rangle =1$;
in particular we aimed at transferring a spin-up state from one end of
the chain to the opposite end, 
or in other words to transform the state $|\psi _i\rangle
=|10...0\rangle$ into the state  
$|\psi _G\rangle =|0...01\rangle$, with $0$ ($1$) corresponding to the
$n$th spin pointing  
in the down (up) direction along the $z$-axis. We employed CRAB to
optimize the two control parameters,  
$\Gamma _1(t)=d(t)$ and $\Gamma _2(t)=C(t)$; as in the previous section,
we set the total evolution time above the quantum speed limit
threshold at the value $T=2T_{\rm QSL}$, 
where for the latter we used the estimate made in
Refs~[\onlinecite{Caneva_PRL09,Murphy_PRA10}]. 
The optimization has been performed by keeping $\vec{\omega}_1,\vec{\omega}_2$ fixed 
(in particular ${\omega}_{1k}={\omega}_{2k}=2k\pi/T$ for $k=1,...,N_c$)
and minimizing the infidelity with respect to
$\vec{A}_1,\vec{B}_1,\vec{A}_2,\vec{B}_2$,  
where the index $1$ and $2$ refer to $d(t)$ and $C(t)$ respectively.

\begin{figure}[t]
\epsfig{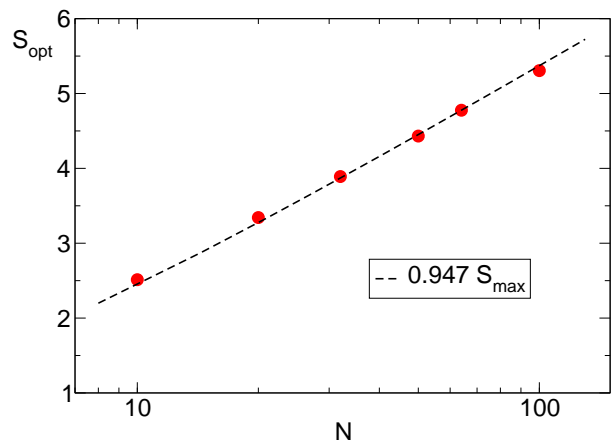}
\caption{(Color online) Entanglement entropy saturation value as a
  function of the size (red circles) and the function
  $A \log _2(N/2+1)$  (dashed line). A fit gives $A=0.947$.} 
\label{lmg_crab_ent_entropy_vs_size_fig}
\end{figure}

The results of our simulations for the state transfer along the chain are summarized 
in Fig.~\ref{crab_chain_inf_vs_n_fig} and Fig.~\ref{crab_chain_inf_vs_npar_fig}.
In Fig.~\ref{crab_chain_inf_vs_n_fig} we show the infidelity as a
function of the size before the optimization (squares), for a constant $C(t)$ and $d(t)=t/T$, and after the
optimization with CRAB (circles): for each size considered we were able to reach 
an infidelity below the value $10^{-4}$ starting from an initial 
infidelity of order $1$.
In Fig.~\ref{crab_chain_inf_vs_npar_fig} we plot the infidelity as a function of 
the number of parameters employed in the minimization procedure; in this case, 
unlike for the LMG model in Fig~\ref{crab_lmg_inf_vs_npar_fig}, the
data show a strong dependence on the size. We interpreted this
behavior as a consequence of the structure  
of the problem. Considering the particular transfer mechanism, in
which the information 
moves step by step from one site to the next one, we expect the
optimal pulse to be able to modulate the magnetic field
around each spin; this occurs only when the spectrum of the pulse 
involves frequencies
of the order of the inverse of the time spent on a generic site $n$, i.e. 
$\omega\sim 2\pi/(T/N)=N\omega _1$. As a test, in the inset of
Fig~\ref{crab_lmg_inf_vs_npar_fig}  
we plotted the infidelity as a function of the number of parameters 
divided by the size;
the good agreement of the rescaled data confirms our expectation.

%
\section{Entanglement entropy maximization}\label{lmg_entanglement:sec}
%
Among its various applications, OC can be exploited for 
entanglement production~\cite{Wang_PRA10, Platzer_PRL10}. 
Here we employ the CRAB technique in the LMG model  
to maximize the von Neumann entropy 
$S_{L,N}=-{\rm Tr}(\rho _{L,N}\log _2\rho _{L,N})$ associated
to the reduced density matrix $\rho _{L,N}$ of a block of $L$ spins
out of the total number $N$ at a given final time $T$, which gives a
measure of the entanglement present between two bipartitions of a 
quantum systems. 
As seen in Sec.~\ref{lmg_crab:sec} due to the symmetry of the
Hamiltonian $[H,\mathcal{S}^2]=0$,  
the dynamics is restricted to subspaces with fixed total angular
momentum; in particular assuming  
as initial state the ground state of the system, we have
$\mathcal{S}=N/2$. The Dicke states  
$|\mathcal{S}=N/2,\mathcal{S}^z \rangle$ with
$\mathcal{S}^z=-N/2,...,N/2$ provide a convenient  
basis spanning the subspace accessible through the dynamics. 
Indeed the entanglement entropy $S_{L,N}$
can be easily evaluated noticing that, since the maximum value of the total spin
can be achieved only with maximum value of the spin in each bipartition, 
the following decomposition holds~\cite{Latorre_PRA05,Caneva_PRB08}:
\begin{eqnarray}
 |N/2,n\rangle &=&\sum _{l=0}^{L} p_{l,n}^{1/2}|L/2,l-L/2\rangle\\\nonumber
&\otimes & |(N-L)/2,n-l-(N-L)/2\rangle,
\end{eqnarray}
where $n$ and $l$ correspond respectively to the number of up spins in
the whole system and in the block of size $L$, and 
$p_{l,n}=L!(N-L)!n!(N-n)!/[l!(L-l)!(n-l)!(N_L-n+l)!N!]$.
Expressing the evolved state $|\psi(T)\rangle$ in the Dicke state
basis and using the previous 
decomposition, it is immediate to evaluate $S_{L,N}(T)$.\\
In our simulations we considered a system equally bipartite,
i.e. $L=N/2$, and we took as starting  
state the ground state of the LMG Hamiltonian 
at $\Gamma \gg 1$, in which all the spins are polarized along the
positive $z$ direction, so that the state factorizes 
and the entanglement entropy vanishes, see 
Fig.~\ref{lmg_crab_ent_vs_T_fig}. Then we performed the optimization
with CRAB, modulating the field according to Eq.~(\ref{field_crab:eq})
and using as a cost function Eq.~(\ref{ent_crab:eq}). 
The behavior of entanglement entropy after the optimization $S_{opt}(T)$  
for different values of the total evolution time $T$ is shown in
Fig.~\ref{lmg_crab_ent_vs_T_fig}: after a short transient 
of linear growth, $S_{opt}(T)$ reaches a
saturation value growing with the size, as expected.
It is interesting to notice that such a behavior closely resembles the
features observed in one-dimensional systems after a sudden 
quench~\cite{Calabrese_JSTAT05}, although here we are dealing with 
a fully connected model~\cite{Dur_PRL05,Latorre_JPA09}.
In Fig.~\ref{lmg_crab_ent_entropy_vs_size_fig} we plotted the saturation value reached
with the optimization as a function of the size $N$; comparing our data
with the maximum possible value of the von Neumann entropy for a
subsystem of $L=N/2$ spins (described by a Hilbert space of dimension
$N/2+1$) $S_{max}=\log _2(N/2+1)$, we obtain almost the maximal
possible amount of entanglement, $S_{opt}/S_{max}\sim
0.95$. 

\section{Linear vs optimal driving}\label{lin_antiad:sec}
%
\begin{figure}[t]
\epsfig{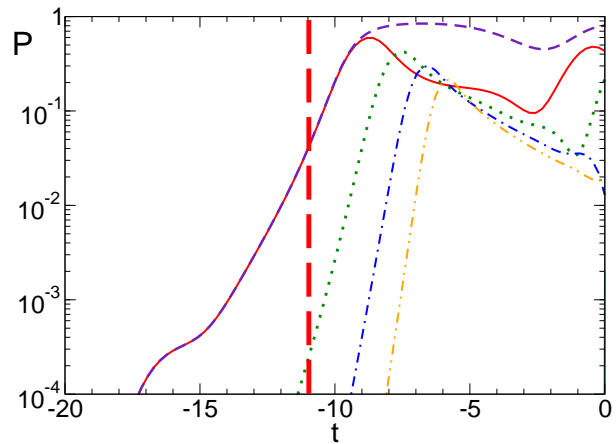}
\caption{(Color online) Instantaneous excitation probabilities $P_i$
of the $i-$th  
excited level ($P_1$ continuous, $P_2$ dotted, $P_3$ dot-dashed, $P_4$ dot-dot-dashed,
$P_{25}$ dash-dash-dotted line)
and total excitation probability
$P_{tot}=\sum_{i=1}^N P_i $ (dashed purple line)
in the LMG model with $N=50$ for an evolution induced with a
driving field linear in time, $\Gamma (t)\propto -t/T$, $T/T_{QSL}=2$,
$N=50$. The thick red
dashed line signals the crossing of the critical point.
The time is measured in units of $J^{-1}$.}
\label{lmg_lin_inst_prob_fig}
\end{figure}
\begin{figure}[t]
\epsfig{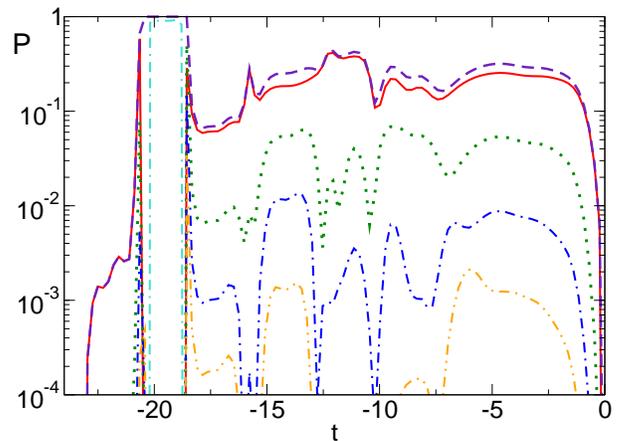}
\caption{(Color online) Optimal instantaneous excitation probabilities $P_i$ 
in the LMG model for $T/T_{QSL}=2$,$N=50$. Codes are the same
as for Fig.~\ref{lmg_lin_inst_prob_fig}. The time is measured in units of $J^{-1}$.
}
\label{lmg_crab_inst_prob_fig}
\end{figure}

%
%
%
In this section we analyze in more detail the features characterizing
the optimal dynamics induced by CRAB. In order to better understand 
the matter, we draw a comparison with 
a simpler non-optimized dynamics, in which the driving field is 
linearly dependent on 
time; in particular we focus the attention on the LMG model. An important
point in the study of the dynamics of a quantum system is usually
represented by the \emph{adiabatic theorem}~\cite{Messiah:book}. 
The latter establishes that a system
initially prepared in its ground state can be driven by a 
time dependent Hamiltonian adiabatically (i.e. without introducing excitations), 
if the time scale of the evolution is much larger than the minimum 
spectral gap, i.e. $T\gg \Delta ^{-1}$.
In critical systems the spectral gap closes at the phase transition,
so that the system gets excited from the instantaneous  
gs while crossing the critical point for any finite-time
evolution~\cite{Sachdev:book}. 
For finite-size systems, the critical gap is not completely closed, 
but it presents a pronounced minimum where the excitation appears,
as shown in Fig~\ref{lmg_lin_inst_prob_fig}: an estimate of the excitations
induced by a linear driving can be obtained by 
Kibble-Zurek theory~\cite{Zener_PRS32,Zurek_PRL05,Caneva_PRB08,Pellegrini_PRB08,Dziarmaga_AP10}.
In the picture we monitored the instantaneous total excitation
probability $P_{tot}$ (dashed line), and the populations of 
lowest levels (different style [color] lines) during the dynamics. 
The evolution starts at large negative times (left) and ends at the
time $t=0$ (right); the critical point is crossed around the time $t=11$
when $\Gamma (t)\sim 1$, see section~\ref{lmg_crab:sec}.  
Far from the critical point the system evolves adiabatically as demonstrated by 
the low total instantaneous excitation probability; notice that before
reaching the critical point the 
total excitation probability coincides with the small excitation of only the
first level (red continuous line).  
In a restricted region around the critical point ($-15<t<-10$)
the total excitation probability jumps to values of order $1$ and does not 
change significantly any more. Notice that in the final
part of the evolution more levels get populated, as shown by the difference 
between the instantaneous infidelity and the excitation probability of
the first level. At the final time $t=T$ the excitation probability is
equal to the infidelity of the process, i.e. $P_{tot}(T)= f_1$.
We then optimize the final infidelity, and the correspondent plot for the optimal 
evolution is reported in Fig~\ref{lmg_crab_inst_prob_fig}. 
The scenario in this case is completely different:
the system is excited at the very beginning of the dynamics and remains excited
for the most part of the evolution until
close to the end, when the infidelity drops abruptly to zero.
It is interesting to notice that just a few levels are excited, as
demonstrated by the small difference between the total excitation
probability (dashed line) and the excitation probability 
of the first level (red continuous line). This result is in agreement with
previous findings where the authors showed that this kind of dynamics
can be approximated by a two-level system dynamics~\cite{Caneva_10:preprint}. 
The abrupt jump in the probabilities around the time $20$ is due to an 
abrupt (double) change of sign in $\Gamma _{\rm CRAB} (t)$, reversing suddenly 
the order of the levels and transforming the gs in the 
most excited state (dash-dash-dotted [cyan] line),
thus this signature is not due to a 
collective involvement of all the levels but simply to a 
reshuffling of their order. Indeed as shown in the picture for 
$-18<t<0$, with the subsequent change of sign the previous 
order is reestablished. We can then summarize the main features of the optimal 
evolution induced by CRAB in three points: it is strongly non adiabatic; 
it involves just a restricted number of levels although not necessarily close to the
nominal instantaneous ground state and it is such that at the very end
all populations constructively interfere to obtain the desired goal
state. 

%

\section{Conclusions}\label{conclusions:sec}
In this paper we studied in detail the performance of quantum optimal
control through the CRAB optimization~\cite{Doria_10:preprint}.
In particular we focused the attention on three different systems and
different figures of merit, in order to outline the versatility of the method.
We first studied the optimization of state transformations of two
qubits via a controlled coupling. We have shown that the CRAB
optimization is very effective already using only a few optimization 
parameters and the fundamental role that the randomization of the 
function basis plays in increasing the process convergence. 
We then analyzed two many-body quantum systems:
the first one, the LMG model, is the prototype of many-body system with 
long range interaction undergoing a quantum phase transition. The
success of CRAB in this context confirms the possibility of 
controlling complex systems typically studied in
condensed matter, with relatively small resources: due to CRAB unique 
features, only few parameters
($3$ frequencies) are indeed sufficient to obtain excellent results.
The second many-body quantum system studied,
the transfer of information along a spin chain, is a typical problem
studied in quantum information theory: 
the high accuracy achievable through CRAB optimization 
makes it a valuable tool for this kind of applications.
Moreover, due to the simple structure of the optimal pulses, they
may be used to extract information on the typical timescales
involved on the system dynamics, as we did for the information
transfer in spin chains. We stress also that the
exponential dependence of the figures of merit as a function of the
number of parameters found in all cases (see Figs.~\ref{minerr},
~\ref{crab_lmg_inf_vs_npar_fig},~\ref{crab_chain_inf_vs_npar_fig}) 
suggests that in general already a moderate number of optimization
parameters will be sufficient to get huge improvements in the
desired processes. 

Finally, we have shown that with a simple change of the cost function, 
the CRAB optimization can be used to optimize the search of the unknown ground 
state of a Hamiltonian or to generate quantum states satisfying 
desired properties, i.e. high entangled states.
Monitoring the instantaneous excitation probabilities
generated by the optimized process, we have demonstrated the highly
non adiabatic character of the dynamics and the fact that, despite the
complexity of the system under study, just a restricted number 
of excited levels are really populated during the evolution. 
The latter fact justifies the compatibility of CRAB with DMRG-like techniques.
We mention that the CRAB optimization has been applied also to
open quantum systems obtaining interesting results and thus increasing
its possible applications~\cite{caruso11}.

In conclusion, the main features of the CRAB optimization 
--versatility (different constraints, compatibility
  with approximate simulation methods and experiments), fast
  convergence (the final error scales exponentially
  with the number of optimization parameters while the
  number of algorithm iterations linearly) and simplicity 
  (small modification to existing numerical codes for quantum 
  system simulations)-- demonstrate that the CRAB optimization is not only 
an unique solution for many body quantum systems optimal control but
it is a valid alternative also in many different settings where other 
optimal control tools exist~\cite{Brif_NJP10}.

We acknowledge discussions with R.~Fazio, V.~Giovannetti, M.~Murphy,
and G.~Santoro, and support from the EU projects AQUTE, PICC, the
SFB/TRR21 and the BWgrid for computational resources.

\bibliographystyle{apsrev}

\begin{thebibliography}{54}
\expandafter\ifx\csname natexlab\endcsname\relax\def\natexlab#1{#1}\fi
\expandafter\ifx\csname bibnamefont\endcsname\relax
  \def\bibnamefont#1{#1}\fi
\expandafter\ifx\csname bibfnamefont\endcsname\relax
  \def\bibfnamefont#1{#1}\fi
\expandafter\ifx\csname citenamefont\endcsname\relax
  \def\citenamefont#1{#1}\fi
\expandafter\ifx\csname url\endcsname\relax
  \def\url#1{\texttt{#1}}\fi
\expandafter\ifx\csname urlprefix\endcsname\relax\def\urlprefix{URL }\fi
\providecommand{\bibinfo}[2]{#2}
\providecommand{\eprint}[2][]{\url{#2}}

\bibitem{Doria_10:preprint} P.~Doria, T.~Calarco, and S.~Montangero
Phys. Rev. Lett. {\bf 106}, 190501 (2011).


\bibitem[{\citenamefont{Brif et~al.}(2010)\citenamefont{Brif, Chakrabarti, and
  Rabitz}}]{Brif_NJP10}
\bibinfo{author}{\bibfnamefont{C.}~\bibnamefont{Brif}},
  \bibinfo{author}{\bibfnamefont{R.}~\bibnamefont{Chakrabarti}},
  \bibnamefont{and} \bibinfo{author}{\bibfnamefont{H.}~\bibnamefont{Rabitz}},
  \bibinfo{journal}{New J. Phys.} \textbf{\bibinfo{volume}{12}},
  \bibinfo{pages}{075008} (\bibinfo{year}{2010}).

\bibitem[{\citenamefont{Feynman}(1982)}]{Feynman_IJTP82}
\bibinfo{author}{\bibfnamefont{R.}~\bibnamefont{Feynman}},
  \bibinfo{journal}{Int. J. Theor. Phys.} \textbf{\bibinfo{volume}{21}},
  \bibinfo{pages}{467} (\bibinfo{year}{1982}).

\bibitem[{\citenamefont{Lloyd}(2000)}]{Lloyd_NAT00}
\bibinfo{author}{\bibfnamefont{S.}~\bibnamefont{Lloyd}},
  \bibinfo{journal}{Nature} \textbf{\bibinfo{volume}{406}},
  \bibinfo{pages}{1047} (\bibinfo{year}{2000}).

\bibitem[{\citenamefont{Bloch et~al.}(2008)\citenamefont{Bloch, Dalibard, and
  Zwerger}}]{Bloch_RMP08}
\bibinfo{author}{\bibfnamefont{I.}~\bibnamefont{Bloch}},
  \bibinfo{author}{\bibfnamefont{J.}~\bibnamefont{Dalibard}}, \bibnamefont{and}
  \bibinfo{author}{\bibfnamefont{W.}~\bibnamefont{Zwerger}},
  \bibinfo{journal}{Rev. Mod. Phys.} \textbf{\bibinfo{volume}{80}},
  \bibinfo{pages}{885} (\bibinfo{year}{2008}).

\bibitem[{\citenamefont{Roati et~al.}(2008)\citenamefont{Roati, D'Errico,
  Fallani, Fattori, Fort, Zaccanti, Modugno, Modugno, and
  Inguscio}}]{Roati_NAT08}
\bibinfo{author}{\bibfnamefont{G.}~\bibnamefont{Roati}},
  \bibinfo{author}{\bibfnamefont{C.}~\bibnamefont{D'Errico}},
  \bibinfo{author}{\bibfnamefont{L.}~\bibnamefont{Fallani}},
  \bibinfo{author}{\bibfnamefont{M.}~\bibnamefont{Fattori}},
  \bibinfo{author}{\bibfnamefont{C.}~\bibnamefont{Fort}},
  \bibinfo{author}{\bibfnamefont{M.}~\bibnamefont{Zaccanti}},
  \bibinfo{author}{\bibfnamefont{G.}~\bibnamefont{Modugno}},
  \bibinfo{author}{\bibfnamefont{M.}~\bibnamefont{Modugno}}, \bibnamefont{and}
  \bibinfo{author}{\bibfnamefont{M.}~\bibnamefont{Inguscio}},
  \bibinfo{journal}{Nature} \textbf{\bibinfo{volume}{453}},
  \bibinfo{pages}{895} (\bibinfo{year}{2008}).

\bibitem[{\citenamefont{Peirce et~al.}(1988)\citenamefont{Peirce, Dahleh, and
  Rabitz}}]{Peirce_PRA88}
\bibinfo{author}{\bibfnamefont{A.~P.} \bibnamefont{Peirce}},
  \bibinfo{author}{\bibfnamefont{M.~A.} \bibnamefont{Dahleh}},
  \bibnamefont{and} \bibinfo{author}{\bibfnamefont{H.}~\bibnamefont{Rabitz}},
  \bibinfo{journal}{Phys. Rev. A} \textbf{\bibinfo{volume}{37}},
  \bibinfo{pages}{4950} (\bibinfo{year}{1988}).

\bibitem[{\citenamefont{Calarco et~al.}(2004)\citenamefont{Calarco, Dorner,
  Julienne, Williams, and Zoller}}]{Calarco_PRA04}
\bibinfo{author}{\bibfnamefont{T.}~\bibnamefont{Calarco}},
  \bibinfo{author}{\bibfnamefont{U.}~\bibnamefont{Dorner}},
  \bibinfo{author}{\bibfnamefont{P.}~\bibnamefont{Julienne}},
  \bibinfo{author}{\bibfnamefont{C.~J.} \bibnamefont{Williams}},
  \bibnamefont{and} \bibinfo{author}{\bibfnamefont{P.}~\bibnamefont{Zoller}},
  \bibinfo{journal}{Phys. Rev. A} \textbf{\bibinfo{volume}{70}},
  \bibinfo{pages}{012306} (\bibinfo{year}{2004}).

\bibitem[{\citenamefont{Khaneja et~al.}(2005)\citenamefont{Khaneja, Reiss,
  Kehlet, Schulte-Herbruggen, and Glaser}}]{Khaneja_JMR05}
\bibinfo{author}{\bibfnamefont{N.}~\bibnamefont{Khaneja}},
  \bibinfo{author}{\bibfnamefont{T.}~\bibnamefont{Reiss}},
  \bibinfo{author}{\bibfnamefont{C.}~\bibnamefont{Kehlet}},
  \bibinfo{author}{\bibfnamefont{T.}~\bibnamefont{Schulte-Herbruggen}},
  \bibnamefont{and} \bibinfo{author}{\bibfnamefont{S.~G.}
  \bibnamefont{Glaser}}, \bibinfo{journal}{J. Magn. Res.}
  \textbf{\bibinfo{volume}{172}}, \bibinfo{pages}{296} (\bibinfo{year}{2005}).

\bibitem[{\citenamefont{Bose}(2003)}]{Bose_PRL03}
\bibinfo{author}{\bibfnamefont{S.}~\bibnamefont{Bose}}, \bibinfo{journal}{Phys.
  Rev. Lett.} \textbf{\bibinfo{volume}{91}}, \bibinfo{pages}{207901}
  (\bibinfo{year}{2003}).

\bibitem[{\citenamefont{Montangero et~al.}(2007)\citenamefont{Montangero,
  Calarco, and Fazio}}]{Montangero_PRL07}
\bibinfo{author}{\bibfnamefont{S.}~\bibnamefont{Montangero}},
  \bibinfo{author}{\bibfnamefont{T.}~\bibnamefont{Calarco}}, \bibnamefont{and}
  \bibinfo{author}{\bibfnamefont{R.}~\bibnamefont{Fazio}},
  \bibinfo{journal}{Phys. Rev. Lett.} \textbf{\bibinfo{volume}{99}},
  \bibinfo{pages}{170501} (\bibinfo{year}{2007}).

\bibitem[{\citenamefont{Krotov}(1996)}]{Krotov:book}
\bibinfo{author}{\bibfnamefont{V.~F.} \bibnamefont{Krotov}},
  \emph{\bibinfo{title}{Global Methods in Optimal Control Theory}}
  (\bibinfo{publisher}{Marcel Dekker}, \bibinfo{address}{New York},
  \bibinfo{year}{1996}).

\bibitem[{\citenamefont{Carlini et~al.}(2006)\citenamefont{Carlini, Hosoya,
  Koike, and Okudaira}}]{Carlini_PRL06}
\bibinfo{author}{\bibfnamefont{A.}~\bibnamefont{Carlini}},
  \bibinfo{author}{\bibfnamefont{A.}~\bibnamefont{Hosoya}},
  \bibinfo{author}{\bibfnamefont{T.}~\bibnamefont{Koike}}, \bibnamefont{and}
  \bibinfo{author}{\bibfnamefont{Y.}~\bibnamefont{Okudaira}},
  \bibinfo{journal}{Phys. Rev. Lett.} \textbf{\bibinfo{volume}{96}},
  \bibinfo{pages}{060503} (\bibinfo{year}{2006}).

\bibitem[{\citenamefont{Rezakhani et~al.}(2009)\citenamefont{Rezakhani, Kuo,
  Hamma, Lidar, and Zanardi}}]{Rezakhani_PRL09}
\bibinfo{author}{\bibfnamefont{A.~T.} \bibnamefont{Rezakhani}},
  \bibinfo{author}{\bibfnamefont{W.-J.} \bibnamefont{Kuo}},
  \bibinfo{author}{\bibfnamefont{A.}~\bibnamefont{Hamma}},
  \bibinfo{author}{\bibfnamefont{D.~A.} \bibnamefont{Lidar}}, \bibnamefont{and}
  \bibinfo{author}{\bibfnamefont{P.}~\bibnamefont{Zanardi}},
  \bibinfo{journal}{Phys. Rev. Lett.} \textbf{\bibinfo{volume}{103}},
  \bibinfo{pages}{080502} (\bibinfo{year}{2009}).

\bibitem[{\citenamefont{Verstraete et~al.}(2009)\citenamefont{Verstraete, Wolf,
  and Cirac}}]{Verstraete_NAT10}
\bibinfo{author}{\bibfnamefont{F.}~\bibnamefont{Verstraete}},
  \bibinfo{author}{\bibfnamefont{M.~M.} \bibnamefont{Wolf}}, \bibnamefont{and}
  \bibinfo{author}{\bibfnamefont{J.~I.} \bibnamefont{Cirac}},
  \bibinfo{journal}{Nat. Phys.} \textbf{\bibinfo{volume}{5}},
  \bibinfo{pages}{633} (\bibinfo{year}{2009}).

\bibitem[{\citenamefont{{Sp\"orl} et~al.}(2007)\citenamefont{{Sp\"orl},
  Schulte-{Herbr\"uggen}, Glaser, Bergholm, Storcz, Ferber, and
  Wilhelm}}]{Sporl_PRA07}
\bibinfo{author}{\bibfnamefont{A.}~\bibnamefont{{Sp\"orl}}},
  \bibinfo{author}{\bibfnamefont{T.}~\bibnamefont{Schulte-{Herbr\"uggen}}},
  \bibinfo{author}{\bibfnamefont{S.~J.}~\bibnamefont{Glaser}},
  \bibinfo{author}{\bibfnamefont{V.}~\bibnamefont{Bergholm}},
  \bibinfo{author}{\bibfnamefont{M.~J.}~\bibnamefont{Storcz}},
  \bibinfo{author}{\bibfnamefont{J.}~\bibnamefont{Ferber}}, \bibnamefont{and}
  \bibinfo{author}{\bibfnamefont{F.~K.}~\bibnamefont{Wilhelm}},
  \bibinfo{journal}{Phys. Rev. A} \textbf{\bibinfo{volume}{75}},
  \bibinfo{pages}{012302} (\bibinfo{year}{2007}).

\bibitem[{\citenamefont{{Schollw\"ock}}(2005)}]{Schollwock_RMP06}
\bibinfo{author}{\bibfnamefont{U.}~\bibnamefont{{Schollw\"ock}}},
  \bibinfo{journal}{Rev. Mod. Phys.} \textbf{\bibinfo{volume}{77}},
  \bibinfo{pages}{259} (\bibinfo{year}{2005}).

\bibitem[{\citenamefont{Romero-Isart and Garcia-Ripoll}(2007)}]{Romero_PRA07}
\bibinfo{author}{\bibfnamefont{O.}~\bibnamefont{Romero-Isart}}
  \bibnamefont{and}
  \bibinfo{author}{\bibfnamefont{J.~J.}~\bibnamefont{Garcia-Ripoll}},
  \bibinfo{journal}{Phys. Rev. A} \textbf{\bibinfo{volume}{76}},
  \bibinfo{pages}{052304} (\bibinfo{year}{2007}).

\bibitem[{\citenamefont{et~al.}(2009)}]{Jiang_PRA09}
\bibinfo{author}{\bibfnamefont{L.} \bibnamefont{Jiang}},
  \bibinfo{author}{\bibfnamefont{A.~M.}~\bibnamefont{Rey}},
  \bibinfo{author}{\bibfnamefont{O.}~\bibnamefont{Romero-Isart}},
  \bibinfo{author}{\bibfnamefont{J.~J.}~\bibnamefont{Garcia-Ripoll}},
  \bibinfo{author}{\bibfnamefont{A.}~\bibnamefont{Sanpera}},
  \bibnamefont{and} \bibinfo{author}{\bibfnamefont{M.~D.}~\bibnamefont{Lukin}},
  \bibinfo{journal}{Phys. Rev. A} \textbf{\bibinfo{volume}{79}},
  \bibinfo{pages}{022309} (\bibinfo{year}{2009}).

\bibitem[{\citenamefont{Galve et~al.}(2010)\citenamefont{Galve, Zueco, Reuther,
  and Kohler}}]{Galve_EPJ10}
\bibinfo{author}{\bibfnamefont{F.}~\bibnamefont{Galve}},
  \bibinfo{author}{\bibfnamefont{D.}~\bibnamefont{Zueco}},
  \bibinfo{author}{\bibfnamefont{G.}~\bibnamefont{Reuther}}, \bibnamefont{and}
  \bibinfo{author}{\bibfnamefont{S.}~\bibnamefont{Kohler}},
  \bibinfo{journal}{Eur. Phys. Journ. Special Topics}
  \textbf{\bibinfo{volume}{180}}, \bibinfo{pages}{237} (\bibinfo{year}{2010}).

\bibitem[{\citenamefont{Francoa et~al.}(2010)\citenamefont{Francoa,
  Paternostro, and Kim}}]{DiFranco_PRA10}
\bibinfo{author}{\bibfnamefont{C.} \bibnamefont{DiFranco}},
  \bibinfo{author}{\bibfnamefont{M.}~\bibnamefont{Paternostro}},
  \bibnamefont{and} \bibinfo{author}{\bibfnamefont{M.~S.}~\bibnamefont{Kim}},
  \bibinfo{journal}{Phys. Rev. A} \textbf{\bibinfo{volume}{81}},
  \bibinfo{pages}{022319} (\bibinfo{year}{2010}).

\bibitem[{\citenamefont{Li et~al.}(2009)\citenamefont{Li, Ruths, and
  Stefanatos}}]{Jr-ShinLi_JCP09}
\bibinfo{author}{\bibfnamefont{J.-S.} \bibnamefont{Li}},
  \bibinfo{author}{\bibfnamefont{J.}~\bibnamefont{Ruths}}, \bibnamefont{and}
  \bibinfo{author}{\bibfnamefont{D.}~\bibnamefont{Stefanatos}},
  \bibinfo{journal}{J. Chem. Phys.} \textbf{\bibinfo{volume}{131}},
  \bibinfo{pages}{164110} (\bibinfo{year}{2009}).

\bibitem[{\citenamefont{Ruths and Li}(2011)}]{Ruths_11:preprint}
\bibinfo{author}{\bibfnamefont{J.}~\bibnamefont{Ruths}} \bibnamefont{and}
  \bibinfo{author}{\bibfnamefont{J.-S.} \bibnamefont{Li}}
  (\bibinfo{year}{2011}), \eprint{arXiv:1102.3713}.

\bibitem[{\citenamefont{Rabitz et~al.}(2004)\citenamefont{Rabitz, Hsieh, and
  Rosenthal}}]{Rabitz_SCI04}
\bibinfo{author}{\bibfnamefont{H.}~\bibnamefont{Rabitz}},
  \bibinfo{author}{\bibfnamefont{M.}~\bibnamefont{Hsieh}}, \bibnamefont{and}
  \bibinfo{author}{\bibfnamefont{C.}~\bibnamefont{Rosenthal}},
  \bibinfo{journal}{Science} \textbf{\bibinfo{volume}{303}},
  \bibinfo{pages}{1998} (\bibinfo{year}{2004}).

\bibitem[{\citenamefont{et~al.}(2006)}]{Rabitz_PRA06}
\bibinfo{author}{\bibfnamefont{H.} \bibnamefont{Rabitz}},
\bibinfo{author}{\bibfnamefont{T.~S.} \bibnamefont{Ho}},
\bibinfo{author}{\bibfnamefont{M.} \bibnamefont{Hsieh}},
\bibinfo{author}{\bibfnamefont{R.} \bibnamefont{Kosut}}, \bibnamefont{and}
\bibinfo{author}{\bibfnamefont{M.} \bibnamefont{Demiralp}},
  \bibinfo{journal}{Phys. Rev. A} \textbf{\bibinfo{volume}{74}},
  \bibinfo{pages}{012721} (\bibinfo{year}{2006}).

\bibitem[{\citenamefont{Caneva et~al.}(2010)\citenamefont{Caneva, Calarco,
  Fazio, Santoro, and Montangero}}]{Caneva_10:preprint}
\bibinfo{author}{\bibfnamefont{T.}~\bibnamefont{Caneva}},
  \bibinfo{author}{\bibfnamefont{T.}~\bibnamefont{Calarco}},
  \bibinfo{author}{\bibfnamefont{R.}~\bibnamefont{Fazio}},
  \bibinfo{author}{\bibfnamefont{G.~E.} \bibnamefont{Santoro}},
  \bibnamefont{and}
  \bibinfo{author}{\bibfnamefont{S.}~\bibnamefont{Montangero}}
  (\bibinfo{year}{2010}), \eprint{arXiv:1011.6634}.

\bibitem[{\citenamefont{Caneva et~al.}(2009)\citenamefont{Caneva, Murphy,
  Calarco, Fazio, Montangero, Giovannetti, and Santoro}}]{Caneva_PRL09}
\bibinfo{author}{\bibfnamefont{T.}~\bibnamefont{Caneva}},
  \bibinfo{author}{\bibfnamefont{M.}~\bibnamefont{Murphy}},
  \bibinfo{author}{\bibfnamefont{T.}~\bibnamefont{Calarco}},
  \bibinfo{author}{\bibfnamefont{R.}~\bibnamefont{Fazio}},
  \bibinfo{author}{\bibfnamefont{S.}~\bibnamefont{Montangero}},
  \bibinfo{author}{\bibfnamefont{V.}~\bibnamefont{Giovannetti}},
  \bibnamefont{and} \bibinfo{author}{\bibfnamefont{G.~E.}
  \bibnamefont{Santoro}}, \bibinfo{journal}{Phys. Rev. Lett.}
  \textbf{\bibinfo{volume}{103}}, \bibinfo{pages}{240501}
  (\bibinfo{year}{2009}).

\bibitem[{\citenamefont{Press et~al.}(1992)\citenamefont{Press, Teukolsky,
  Vetterling, and Flannery}}]{NumericalRecipes}
\bibinfo{author}{\bibfnamefont{W.~H.} \bibnamefont{Press}},
  \bibinfo{author}{\bibfnamefont{S.~A.} \bibnamefont{Teukolsky}},
  \bibinfo{author}{\bibfnamefont{W.~T.} \bibnamefont{Vetterling}},
  \bibnamefont{and} \bibinfo{author}{\bibfnamefont{B.~P.}
  \bibnamefont{Flannery}}, \emph{\bibinfo{title}{Numerical recipes in {C}: the
  art of scientific computing}} (\bibinfo{publisher}{Cambridge University
  Press}, \bibinfo{year}{1992}), \bibinfo{edition}{2nd} ed.

\bibitem[{\citenamefont{Makhlin et~al.}(2001)\citenamefont{Makhlin, {Sch\"on},
  and Shnirman}}]{Mahklin_RMP01}
\bibinfo{author}{\bibfnamefont{Y.}~\bibnamefont{Makhlin}},
  \bibinfo{author}{\bibfnamefont{G.}~\bibnamefont{{Sch\"on}}},
  \bibnamefont{and} \bibinfo{author}{\bibfnamefont{A.}~\bibnamefont{Shnirman}},
  \bibinfo{journal}{Rev. Mod. Phys.} \textbf{\bibinfo{volume}{73}},
  \bibinfo{pages}{357} (\bibinfo{year}{2001}).

\bibitem[{\citenamefont{Wendin and Shumeiko}(2006)}]{Wendin_06:inc}
\bibinfo{author}{\bibfnamefont{G.}~\bibnamefont{Wendin}} \bibnamefont{and}
  \bibinfo{author}{\bibfnamefont{V.}~\bibnamefont{Shumeiko}}, in
  \emph{\bibinfo{booktitle}{{Handbook of Theoretical and Computational
  Nanotechnology}}}, edited by
  \bibinfo{editor}{\bibfnamefont{M.}~\bibnamefont{Rieth}} \bibnamefont{and}
  \bibinfo{editor}{\bibfnamefont{W.}~\bibnamefont{Schommers}}
  (\bibinfo{publisher}{American Scientific Publishers}, \bibinfo{year}{2006}),
  p. \bibinfo{pages}{Ch. 12}, \eprint{cond-mat/0508729}.

\bibitem[{\citenamefont{Lipkin et~al.}(1965)\citenamefont{Lipkin, Meshkov, and
  Glick}}]{Lipkin_NP65}
\bibinfo{author}{\bibfnamefont{H.~J.} \bibnamefont{Lipkin}},
  \bibinfo{author}{\bibfnamefont{N.}~\bibnamefont{Meshkov}}, \bibnamefont{and}
  \bibinfo{author}{\bibfnamefont{A.~J.} \bibnamefont{Glick}},
  \bibinfo{journal}{Nucl. Phys.} \textbf{\bibinfo{volume}{62}},
  \bibinfo{pages}{188} (\bibinfo{year}{1965}).

\bibitem[{\citenamefont{Botet and Jullien}(1983)}]{Botet_PRB83}
\bibinfo{author}{\bibfnamefont{R.}~\bibnamefont{Botet}} \bibnamefont{and}
  \bibinfo{author}{\bibfnamefont{R.}~\bibnamefont{Jullien}},
  \bibinfo{journal}{Phys. Rev. B} \textbf{\bibinfo{volume}{28}},
  \bibinfo{pages}{3955} (\bibinfo{year}{1983}).

\bibitem[{\citenamefont{Messiah}(1962)}]{Messiah:book}
\bibinfo{author}{\bibfnamefont{A.}~\bibnamefont{Messiah}},
  \emph{\bibinfo{title}{Quantum mechanics}}, vol.~\bibinfo{volume}{2}
  (\bibinfo{publisher}{North-Holland}, \bibinfo{address}{Amsterdam},
  \bibinfo{year}{1962}).

\bibitem[{\citenamefont{Sachdev}(1999)}]{Sachdev:book}
\bibinfo{author}{\bibfnamefont{S.}~\bibnamefont{Sachdev}},
  \emph{\bibinfo{title}{Quantum Phase Transition}}
  (\bibinfo{publisher}{Cambridge University Press}, \bibinfo{year}{1999}).

\bibitem[{\citenamefont{Zurek et~al.}(2005)\citenamefont{Zurek, Dorner, and
  Zoller}}]{Zurek_PRL05}
\bibinfo{author}{\bibfnamefont{W.~H.} \bibnamefont{Zurek}},
  \bibinfo{author}{\bibfnamefont{U.}~\bibnamefont{Dorner}}, \bibnamefont{and}
  \bibinfo{author}{\bibfnamefont{P.}~\bibnamefont{Zoller}},
  \bibinfo{journal}{Phys. Rev. Lett.} \textbf{\bibinfo{volume}{95}},
  \bibinfo{pages}{105701} (\bibinfo{year}{2005}).

\bibitem[{\citenamefont{Polkovnikov and Gritsev}(2008)}]{Polkovnikov_NAT08}
\bibinfo{author}{\bibfnamefont{A.}~\bibnamefont{Polkovnikov}} \bibnamefont{and}
  \bibinfo{author}{\bibfnamefont{V.}~\bibnamefont{Gritsev}},
  \bibinfo{journal}{Nature Physics} \textbf{\bibinfo{volume}{4}},
  \bibinfo{pages}{477} (\bibinfo{year}{2008}).

\bibitem[{\citenamefont{Pellegrini et~al.}(2008)\citenamefont{Pellegrini,
  Montangero, Santoro, and Fazio}}]{Pellegrini_PRB08}
\bibinfo{author}{\bibfnamefont{F.}~\bibnamefont{Pellegrini}},
  \bibinfo{author}{\bibfnamefont{S.}~\bibnamefont{Montangero}},
  \bibinfo{author}{\bibfnamefont{G.~E.}~\bibnamefont{Santoro}}, \bibnamefont{and}
  \bibinfo{author}{\bibfnamefont{R.}~\bibnamefont{Fazio}},
  \bibinfo{journal}{Phys. Rev. B} \textbf{\bibinfo{volume}{77}},
  \bibinfo{pages}{140404(R)} (\bibinfo{year}{2008}).

\bibitem[{\citenamefont{Deng et~al.}(2008)\citenamefont{Deng, Ortiz, and
  Viola}}]{Deng_EPL08}
\bibinfo{author}{\bibfnamefont{S.}~\bibnamefont{Deng}},
  \bibinfo{author}{\bibfnamefont{G.}~\bibnamefont{Ortiz}}, \bibnamefont{and}
  \bibinfo{author}{\bibfnamefont{L.}~\bibnamefont{Viola}},
  \bibinfo{journal}{Europhys. Lett.} \textbf{\bibinfo{volume}{84}},
  \bibinfo{pages}{67008} (\bibinfo{year}{2008}).

\bibitem[{\citenamefont{Grandi and Polkovnikov}(2009)}]{DeGrandi_09:preprint}
\bibinfo{author}{\bibfnamefont{C.~D.} \bibnamefont{Grandi}} \bibnamefont{and}
  \bibinfo{author}{\bibfnamefont{A.}~\bibnamefont{Polkovnikov}}
  (\bibinfo{year}{2009}), \eprint{arXiv:0910.2236}.

\bibitem[{\citenamefont{Divakaran et~al.}(2009)\citenamefont{Divakaran,
  Mukherjee, Dutta, and Sen}}]{Divakaran_09:preprint}
\bibinfo{author}{\bibfnamefont{U.}~\bibnamefont{Divakaran}},
  \bibinfo{author}{\bibfnamefont{V.}~\bibnamefont{Mukherjee}},
  \bibinfo{author}{\bibfnamefont{A.}~\bibnamefont{Dutta}}, \bibnamefont{and}
  \bibinfo{author}{\bibfnamefont{D.}~\bibnamefont{Sen}} (\bibinfo{year}{2009}),
  \eprint{arXiv:0908.4004}.

\bibitem[{\citenamefont{Dziarmaga}(2010)}]{Dziarmaga_AP10}
\bibinfo{author}{\bibfnamefont{J.}~\bibnamefont{Dziarmaga}},
  \bibinfo{journal}{Adv. Phys.} \textbf{\bibinfo{volume}{59}},
  \bibinfo{pages}{1063} (\bibinfo{year}{2010}).

\bibitem[{\citenamefont{Polkovnikov et~al.}(2010)\citenamefont{Polkovnikov,
  Sengupta, Silva, and Vengalattore}}]{Polkovnikov_10:preprint}
\bibinfo{author}{\bibfnamefont{A.}~\bibnamefont{Polkovnikov}},
  \bibinfo{author}{\bibfnamefont{K.}~\bibnamefont{Sengupta}},
  \bibinfo{author}{\bibfnamefont{A.}~\bibnamefont{Silva}}, \bibnamefont{and}
  \bibinfo{author}{\bibfnamefont{M.}~\bibnamefont{Vengalattore}}
  (\bibinfo{year}{2010}), \eprint{arXiv:1007.5331}.

\bibitem[{\citenamefont{Caneva et~al.}(2008)\citenamefont{Caneva, Fazio, and
  Santoro}}]{Caneva_PRB08}
\bibinfo{author}{\bibfnamefont{T.}~\bibnamefont{Caneva}},
  \bibinfo{author}{\bibfnamefont{R.}~\bibnamefont{Fazio}}, \bibnamefont{and}
  \bibinfo{author}{\bibfnamefont{G.~E.} \bibnamefont{Santoro}},
  \bibinfo{journal}{Phys. Rev. B} \textbf{\bibinfo{volume}{78}},
  \bibinfo{pages}{104426} (\bibinfo{year}{2008}).

\bibitem[{\citenamefont{Giovannetti et~al.}(2003)\citenamefont{Giovannetti,
  Lloyd, and Maccone}}]{Giovannetti_PRA03}
\bibinfo{author}{\bibfnamefont{V.}~\bibnamefont{Giovannetti}},
  \bibinfo{author}{\bibfnamefont{S.}~\bibnamefont{Lloyd}}, \bibnamefont{and}
  \bibinfo{author}{\bibfnamefont{L.}~\bibnamefont{Maccone}},
  \bibinfo{journal}{Phys. Rev. A} \textbf{\bibinfo{volume}{67}},
  \bibinfo{pages}{052109} (\bibinfo{year}{2003}).

\bibitem[{\citenamefont{Murphy et~al.}(2010)\citenamefont{Murphy, Montangero,
  Giovannetti, and Calarco}}]{Murphy_PRA10}
\bibinfo{author}{\bibfnamefont{M.}~\bibnamefont{Murphy}},
  \bibinfo{author}{\bibfnamefont{S.}~\bibnamefont{Montangero}},
  \bibinfo{author}{\bibfnamefont{V.}~\bibnamefont{Giovannetti}},
  \bibnamefont{and} \bibinfo{author}{\bibfnamefont{T.}~\bibnamefont{Calarco}},
  \bibinfo{journal}{Phys. Rev. A} \textbf{\bibinfo{volume}{82}},
  \bibinfo{pages}{022318} (\bibinfo{year}{2010}).

\bibitem[{\citenamefont{Balachandran and Gong}(2008)}]{Balachandran_PRA08}
\bibinfo{author}{\bibfnamefont{V.}~\bibnamefont{Balachandran}}
  \bibnamefont{and} \bibinfo{author}{\bibfnamefont{J.}~\bibnamefont{Gong}},
  \bibinfo{journal}{Phys. Rev. A} \textbf{\bibinfo{volume}{77}},
  \bibinfo{pages}{012303} (\bibinfo{year}{2008}).

\bibitem[{\citenamefont{Wang et~al.}(2010)\citenamefont{Wang, Bayat, Schirmer,
  and Bose}}]{Wang_PRA10}
\bibinfo{author}{\bibfnamefont{X.}~\bibnamefont{Wang}},
  \bibinfo{author}{\bibfnamefont{A.}~\bibnamefont{Bayat}},
  \bibinfo{author}{\bibfnamefont{S.~G.} \bibnamefont{Schirmer}},
  \bibnamefont{and} \bibinfo{author}{\bibfnamefont{S.}~\bibnamefont{Bose}},
  \bibinfo{journal}{Phys. Rev. A} \textbf{\bibinfo{volume}{81}},
  \bibinfo{pages}{032312} (\bibinfo{year}{2010}).

\bibitem[{\citenamefont{Platzer et~al.}(2010)\citenamefont{Platzer, Mintert,
  and Buchleitner}}]{Platzer_PRL10}
\bibinfo{author}{\bibfnamefont{F.}~\bibnamefont{Platzer}},
  \bibinfo{author}{\bibfnamefont{F.}~\bibnamefont{Mintert}}, \bibnamefont{and}
  \bibinfo{author}{\bibfnamefont{A.}~\bibnamefont{Buchleitner}},
  \bibinfo{journal}{Phys. Rev. Lett.} \textbf{\bibinfo{volume}{105}},
  \bibinfo{pages}{020501} (\bibinfo{year}{2010}).

\bibitem[{\citenamefont{Latorre et~al.}(2005)\citenamefont{Latorre, Orus, Rico,
  and Vidal}}]{Latorre_PRA05}
\bibinfo{author}{\bibfnamefont{J.~I.} \bibnamefont{Latorre}},
  \bibinfo{author}{\bibfnamefont{R.}~\bibnamefont{Orus}},
  \bibinfo{author}{\bibfnamefont{E.}~\bibnamefont{Rico}}, \bibnamefont{and}
  \bibinfo{author}{\bibfnamefont{J.}~\bibnamefont{Vidal}},
  \bibinfo{journal}{Phys. Rev. A} \textbf{\bibinfo{volume}{71}},
  \bibinfo{pages}{064101} (\bibinfo{year}{2005}).

\bibitem[{\citenamefont{Calabrese and Cardy}(2005)}]{Calabrese_JSTAT05}
\bibinfo{author}{\bibfnamefont{P.}~\bibnamefont{Calabrese}} \bibnamefont{and}
  \bibinfo{author}{\bibfnamefont{J.}~\bibnamefont{Cardy}}, \bibinfo{journal}{J.
  Stat. Mech.} p. \bibinfo{pages}{P04010} (\bibinfo{year}{2005}).

\bibitem[{\citenamefont{Dur et~al.}(2005)\citenamefont{Dur, Hartmann, Hein,
  Lewenstein, and Briegel}}]{Dur_PRL05}
\bibinfo{author}{\bibfnamefont{W.}~\bibnamefont{Dur}},
  \bibinfo{author}{\bibfnamefont{L.}~\bibnamefont{Hartmann}},
  \bibinfo{author}{\bibfnamefont{M.}~\bibnamefont{Hein}},
  \bibinfo{author}{\bibfnamefont{M.}~\bibnamefont{Lewenstein}},
  \bibnamefont{and} \bibinfo{author}{\bibfnamefont{H.-J.}
  \bibnamefont{Briegel}}, \bibinfo{journal}{Phys. Rev. Lett.}
  \textbf{\bibinfo{volume}{94}}, \bibinfo{pages}{097203}
  (\bibinfo{year}{2005}).

\bibitem[{\citenamefont{Latorre and Riera}(2009)}]{Latorre_JPA09}
\bibinfo{author}{\bibfnamefont{J.~I.} \bibnamefont{Latorre}} \bibnamefont{and}
  \bibinfo{author}{\bibfnamefont{A.}~\bibnamefont{Riera}}, \bibinfo{journal}{J.
  Phys. A: Math. Theor.} \textbf{\bibinfo{volume}{42}}, \bibinfo{pages}{504002}
  (\bibinfo{year}{2009}).

\bibitem[{\citenamefont{Zener}(1932)}]{Zener_PRS32}
\bibinfo{author}{\bibfnamefont{C.}~\bibnamefont{Zener}},
  \bibinfo{journal}{Proc. Royal Soc. A} \textbf{\bibinfo{volume}{137}},
  \bibinfo{pages}{696} (\bibinfo{year}{1932}).

\bibitem[{car()}]{caruso11}
\bibinfo{note}{F.~Caruso, S.~Montangero, T.~Calarco, S.F.~Huelga, and
  M.B.~Plenio, arXiv:1103.0929.}

\end{thebibliography}

\end{document}